\documentclass[10pt,conference,switch]{IEEEtran}
\IEEEoverridecommandlockouts
\usepackage{amsmath,amssymb,amsfonts}
\usepackage{algorithmic}
\usepackage{graphicx}
\usepackage{textcomp}
\usepackage[table]{xcolor}
\def\BibTeX{{\rm B\kern-.05em{\sc i\kern-.025em b}\kern-.08em
    T\kern-.1667em\lower.7ex\hbox{E}\kern-.125emX}}

\usepackage{enumitem}

\usepackage{todonotes}
\usepackage[numbers, sort&compress]{natbib}
\usepackage{url}
\usepackage{booktabs}
\usepackage{caption}
\usepackage{multirow}
\usepackage{censor}
\usepackage{hyperref}
\usepackage{soul}
\newcommand{\code}[1]{%
  \begingroup
  \sethlcolor{lightgray}%
  \hl{#1}%
  \endgroup
}
\usepackage{subcaption}
\usepackage{xcolor}
\newcommand{\email}[1]{\href{mailto:#1}{#1}}

\makeatletter
\newcommand{\linebreakand}{%
  \end{@IEEEauthorhalign}
  \hfill\mbox{}\par
  \mbox{}\hfill\begin{@IEEEauthorhalign}
}
\makeatother

\usepackage[frozencache ,cachedir=.]{minted} %finalizecache

\definecolor{codegreen}{rgb}{0.0, 0.5, 0.0}

\setminted[C]{ %
    linenos=false,             % Line numbers
    autogobble=true,          % Automatically remove common white space
    frame=lines,
    framesep=2mm,
    fontsize=\footnotesize
    }

\begin{document}

\title{Extending Source Code Pre-Trained Language Models to Summarise Decompiled Binaries\\
% \thanks{Applicable funding}
}

\author{
\IEEEauthorblockN{Ali Al-Kaswan}
\IEEEauthorblockA{\textit{Delft University of Technology} \\
Delft, The Netherlands \\
\email{a.al-kaswan@tudelft.nl}} 
\and
\IEEEauthorblockN{Toufique Ahmed}
\IEEEauthorblockA{\textit{University of California, Davis} \\
Davis, California, USA \\
\email{tfahmed@ucdavis.edu}}
\and
\IEEEauthorblockN{Maliheh Izadi}
\IEEEauthorblockA{\textit{Delft University of Technology} \\
Delft, The Netherlands \\
\email{m.izadi@tudelft.nl}}
\linebreakand
\IEEEauthorblockN{Anand Ashok Sawant}
\IEEEauthorblockA{\textit{University of California, Davis} \\
Davis, California, USA \\
\email{asawant@ucdavis.edu}}
\and 
\IEEEauthorblockN{Premkumar Devanbu}
\IEEEauthorblockA{\textit{University of California, Davis} \\
Davis, California, USA \\
\email{ptdevanbu@ucdavis.edu}}
\and
\IEEEauthorblockN{Arie van Deursen}
\IEEEauthorblockA{\textit{Delft University of Technology} \\
Delft, The Netherlands \\
\email{arie.vandeursen@tudelft.nl}}
}

\maketitle

\begin{abstract}
\emph{Binary reverse engineering} is used to understand and analyse programs for which the source code is unavailable. 
Decompilers can help, transforming opaque binaries into a more readable source code-like representation. 
Still, reverse engineering is difficult and costly, 
involving considering effort in labelling code
with helpful summaries. 
%Current status and challenges
While the automated summarisation of decompiled code can help reverse engineers understand and analyse binaries, current work mainly focuses on summarising source code, and no suitable dataset exists for this task.
% proposed
In this work, 
we extend large pre-trained language models of source code 
to summarise de-compiled binary functions.
Furthermore, we investigate 
the impact of \textit{input} and \textit{data properties} 
on the performance of such models.
Our approach consists of two main components; 
the data and the model.
We first build CAPYBARA, 
a dataset of 214K decompiled function-documentation pairs 
across various compiler optimisations.
We extend CAPYBARA further by removing identifiers, and deduplicating the data. 
Next, we fine-tune the CodeT5 base model with CAPYBARA to create BinT5. 
% results
% Experimental results show that 
BinT5 achieves the state-of-the-art BLEU-4 score of 60.83, 58.82 and, 44.21 
for summarising source, decompiled, and obfuscated decompiled code, respectively. 
This indicates that these models can be extended to decompiled binaries successfully.
% Further testing reveals that duplication only decreases the BLEU-4 score 24\% 28\% and 43\%. 
Finally, we found that the performance of BinT5 is not heavily dependent on the dataset size
and compiler optimisation level.
We recommend future research to further investigate 
transferring knowledge 
when working with less expressive input formats
such as stripped binaries.
\end{abstract}

\section{Introduction}
\label{introduction}
%Intro binary reverse engineering
Reverse engineering binary programs has many applications, in particular, software security~\cite{reverseEngineerProcess}. Binary reverse engineering is a hard task, requiring highly skilled reverse engineers~\cite{reverseEngineerProcess, Nero}. Disassemblers and decompilers can help in this process. Disassemblers transform the binary into a low-level intermediate representation, and decompilers lift the representation to a high-level programming language-like representation. But the output of decompilers is still difficult to read and understand~\cite{reverseEngineerProcess, TypeInferenceSurvey}. Much of the work that goes into reverse engineering a binary is spent labelling functions with semantic descriptions~\cite{reverseEngineerProcess}.  Current approaches~\cite{CATI,Debin,Dire, FUNCRE, symLM, SnowWhite, Neutron} mainly focus on recovering aspects lost in the compilation and decompilation process, such as names and types. 
%This fails to address the need for methods to increase the comprehensibility of decompiled code, since they do not provide a high-level view of the program. 
Existing works fail to address the inherent difficulties in binary code comprehensibility, namely, the need for a high-level overview of the code.

%Code sum
For source code, methods exist to automatically generate summaries from code~\cite{sourceCodeSumSurvey, towardsCodeSum}. Source code summarisation is used to automatically generate short natural language descriptions of code, which support program comprehension and aid maintenance~\cite{towardsCodeSum, evaluationSummarization}. While these methods have been successfully applied to programming languages such as Python, Java and PHP~\cite{CodeT5, CodeBERT, PolyglotCodeBERT}, using pre-trained language models~\cite{CodeT5,CodeBERT,PolyglotCodeBERT}, none of these methods has been applied to the relatively syntactically-poor output of decompilers (see Figures ~\ref{fig:sourceExample} and ~\ref{fig:decomExample}). Being able to quickly determine the context and application of a function, can save valuable analysis time, and greatly benefit reverse engineers. Function and variable names alone, are inadequate representations of the source code~\cite{towardsCodeSum}, which is why having descriptive summaries of binaries is desirable.

%Dual channels
Following~\cite{dualChannel}, source code can be described as having two information channels: the algorithmic channel and the natural language channel. The algorithmic channel specifies the execution of a program (semantics), while the natural language channel explains the purpose and context of the program to humans~\cite{dualChannel}. The natural channel includes function and variable names, code comments and the specific human-readable structure of programs. Processors only consider the algorithmic channel to execute a program, while humans use both the algorithmic channel and the natural channel to understand a piece of code~\cite{dualChannel}. Furthermore, code is very regular and predictable, even more so than natural languages~\cite{naturalnessCode}. 
%Use dual channel theory to explain how challening binary reverse engineering is
The compilation process, which transforms readable code into executable binaries, removes much of the information contained in the natural channel. Especially stripped binaries --- binaries of which the symbol table is removed --- are challenging, since they have almost no identifiers at all as can be observed in Figure~\ref{fig:stripExample}.

%State goal of paper
The goal of this paper is
to advance the field of binary reverse engineering by
exploring the application of code summarisation to decompiled binaries
by taking advantage of source code pre-trained language models.

%Need for a dataset + Capybara
However, there exists no dataset of aligned binaries 
and source code summaries since this is a new and unexplored task. As pointed out by~\citeauthor{recommend_summarization}, the lack of standardised datasets is a major barrier to ongoing research, which we will address for this task~\cite{recommend_summarization}. In this paper, we create a dataset containing pairs of decompiled and stripped-decompiled functions and summaries of these functions. During the creation of this dataset, we conform to the current best practices for dataset construction~\cite{recommend_summarization, Codesearchnet}.

%The model BinT5
We apply this dataset to an existing pre-trained language model using transfer learning, by fine-tuning this pre-trained model on our dataset. For this task, we selected a pre-trained CodeT5 model, which was only trained on source code~\cite{CodeT5}.

%Experiments: Explore impact of decompilation, removing identifiers, stripping, duplication
We perform experiments on this model to explore the impact of decompilation, and the importance of identifiers. Furthermore, we explore the impact of compiler optimisation levels, the dataset size and the level of duplication. 

Our findings are that the decompilation and alignment of stripped functions has a very high failure rate; and the resulting stripped model has low performance. But, we found that the model shows state-of-the-art performance with both decompiled code as well as demi-stripped stripped code, code of which the identifiers were removed after decompilation. Our experiments on data duplication and dataset size further show that these models can be trained with few data, and that while duplicates have a high impact on performance, their presence is not paramount to model performance.

%Our key result
Our key result: \textit{language models pre-trained on source code can be fine-tuned on binaries, opening up a range of new possibilities for the automated analysis of binaries.}

\textbf{To summarise, the main contributions of this paper are:}
\begin{itemize}
    \item CAPYBARA\footnote{CAPYBARA: \url{https://doi.org/10.5281/zenodo.7229809}}, a dataset of \textit{\textbf{C}ombined \textbf{A}ligned decom\textbf{P}iled \textbf{B}inar\textbf{Y} code \textbf{A}nd \textbf{R}elated \textbf{A}nnotations}. A novel dataset of aligned, C, decompiled, stripped-decompiled and demi-stripped summary pairs\footnote{Decompiled code with strip-like obfuscation applied} (Section~\ref{dataset}); 
    \item BinT5\footnote{BinT5: \url{https://doi.org/10.5281/zenodo.7229913}}, a \textbf{Bin}ary summarisation Code\textbf{T5} model, a simple and straightforward adaptation of a source code trained code summarisation model to decompiled code using CAPYBARA (Section~\ref{model});
    \item An empirical investigation on the impact of the properties of decompiled code and the properties of CAPYBARA (Sections~\ref{expSetup} and ~\ref{results});
\end{itemize}
The materials, including the processed and raw data, the trained model checkpoints and steps to replicate our experiments, are openly available in our replication package\footnote{Replication package: \url{https://github.com/AISE-TUDelft/Capybara-BinT5}}.

\section{Background}
\label{background}
In this section, we introduce the background of compilers, binary reverse engineering, transfer learning and the code summarisation task.

\subsection{Compilers and Optimisation Levels}
Compilers are programs that convert source code from one programming language to another, but generally, and in the context of this work, the term is used to refer to programs that translate high-level code, like C, to a lower-level language such as machine code or bytecode. For our work, we focus on the GNU Compiler Collection (GCC)\footnote{GCC: \url{https://gcc.gnu.org/}} and Clang/LLVM (Clang)\footnote{Clang: \url{https://clang.llvm.org/}}.

Compilers feature optimisation levels. Generally, the goal of optimisations is the improvement of runtime performance or program size at the expense of compilation time and the ability to debug~\cite{ColeOptimizationLevel}.

By default, if GCC is invoked without any optimisation options, the program will be compiled with -O0. -O1, -O2 and -O3 incrementally apply more optimisation to the binary at the expense of a higher compilation time~\cite{gccOptimization}. Optimisations can restructure and transform the program in relation to the source code, by changing the control flow or the data of the program~\cite{optimizationObfuscation}. This obfuscation can complicate the reverse engineering process by reducing the accuracy of tools~\cite{optimizationObfuscation}.  

\begin{figure}[h!]
        \centering
        \begin{subfigure}{\linewidth}
            \centering
            \begin{minted}[breaklines,escapeinside=||]{C}
            /**
             * Get the Synchronizing source for an RTP/RTCP Socket
             * |\textcolor{codegreen}{\textbf{@param}}| rs RTP Socket
             * |\textcolor{codegreen}{\textbf{@return}}| Synchronizing source
             */
            uint32_t rtp_sess_ssrc(const struct rtp_sock *rs){
            	return rs ? rs -> enc.ssrc : 0;}
            \end{minted}
            \caption{Source rtp\_sess\_ssrc function}
            \label{fig:sourceExample}
            \hfill
        \end{subfigure}
        \hfill
        \begin{subfigure}{\linewidth}
            \centering
            \begin{minted}[breaklines]{C}
            ulong rtp_sess_ssrc(long param_1){ 
                uint local_14 ; 
                if (param_1 == 0){ 
                    local_14 = 0;
                } else {    
                    local_14 = * (uint *) (param_1 + 4);}
                return (ulong) local_14;
            }
            \end{minted}
            \caption{Decompiled rtp\_sess\_ssrc function}
            \label{fig:decomExample}
            \hfill
        \end{subfigure} 
        \hfill
        \begin{subfigure}{\linewidth}
            \centering
            \begin{minted}[breaklines]{C}
            ulong FUN_00100d30 ( long param_1 ){ 
                uint local_14 ;
                if (param_1 == 0) { 
                    local_14 = 0 ;  
                } else { 
                    local_14 = * (uint *) (param_1 + 4);}
                return ( ulong ) local_14 ;}
            \end{minted}
            \caption{Stripped decompiled rtp\_sess\_ssrc function}
            \label{fig:stripExample}
        \end{subfigure} 
    \caption{Example source, decompiled and stripped code snippet}
\end{figure}

\subsection{Ghidra}
Ghidra\footnote{Ghidra: \url{https://ghidra-sre.org/}} is a free and open-source reverse engineering toolkit developed by the US National Security Agency. Ghidra contains many separate analysis modules that allow a reverse engineer to analyse binaries. Ghidra features a disassembler, which assembles binaries back into an intermediate representation. In the case of x86-x64 binaries like the binaries this work focuses on, the intermediate representation will be the Assembly language. The decompiler, on the other hand, is a processor language-agnostic transformation engine that takes the disassembled code and creates a source code representation, namely pseudo-C. Pseudo-C follows the general language conventions of C, but it cannot be compiled. 

Observe the relatively simple \code{rtp\_sess\_ssrc} function from \code{creytiv/re}\footnote{re: \url{https://github.com/creytiv/re}} shown in Figure \ref{fig:sourceExample}. We compile the project using the -O3 compiler level as defined in the project. We decompile the binaries using Ghidra's decompiler using the standard configuration, the resulting pseudo-code is shown in Figure~\ref{fig:decomExample}. We observe that aside from the function name, almost the entire natural channel has been destroyed by the compilation and decompilation process. The parameter and variable names are gone, any documentation is removed and the relatively simple logic has been unrolled to a much more difficult-to-understand representation. Ghidra also incorrectly labelled many of the variable types and failed to identify the \code{struct} datatype.

Using our trained BinT5 model we can summarise the decompiled code and generate the following summary: \code{Get the source for an RTP/RTCP Socket}. This summary gives us an indication of the purpose of the function. Integrating this generated summary into Ghidra increases the readability of the entire binary. Keep in mind that a reverse engineer has to understand not just this function, but hundreds of different functions in a single binary.

\subsection{Stripping}
Aside from compiling with higher optimisation levels, binaries can also be stripped to obfuscate the underlying code and to resist analysis~\cite{StochFuzz}. Commercial off-the-shelf software is often stripped to reduce the memory and storage footprint of the binaries, and to resist analysis to protect the intellectual property of the creator. Many vulnerable and malicious binaries are, unfortunately, also stripped to resist security analysis and hide their faults~\cite{Debin}.

Unix and Unix-like operating systems include a strip utility. The strip utility removes any operands that are not necessary for the execution of the binary while ensuring that the execution of the binary remains unchanged. The exact implementation and what constitutes unnecessary operands are left to the implementor.\footnote{strip: \url{https://pubs.opengroup.org/onlinepubs/7908799/xcu/strip.html}} The strip utility as implemented in GNU/Linux removes the symbol table from the binary. The symbol table contains each symbol's location, type and name.

Like higher optimisation levels, the use of stripping can greatly complicate the efforts to reverse engineer a binary, as well as reduce the accuracy and effectiveness of reverse engineering tools~\cite{StochFuzz}. 

For example, we compile, strip and decompile the function in Figure~\ref{fig:sourceExample}, and the resulting stripped decompiled function is shown in Figure~\ref{fig:stripExample}. In addition to the details lost by the decompilation process, the stripper removed all symbols, like the function names.

\subsection{Code Summarisation Task:}
Code summarisation (also referred to as source code summarisation) is the task of writing short descriptions from source code, usually a single-sentence summary of the source code. The main use is for software documentation, like the one-sentence JavaDoc description used in Java~\cite{recommend_summarization}. This documentation is important for program comprehension and maintenance. But the process of writing and maintaining these descriptions is a labour-intensive and time-consuming task, which is where the benefits of automating that process arise. Automatic code summarisation is an active and popular research problem in the field of software engineering~\cite{recommend_summarization}.

\subsection{Transformer-based Models}
Transformers were originally proposed by~\citeauthor{Transformers} as a sequence-to-sequence architecture~\cite{Transformers}. Unlike the Recurrent Neural Networks~\cite{RNN} (RNN), the Long Short-Term Memory~\cite{LSTM} (LSTM) variant of RNNs~\cite{RNN} and Convolutional Neural Networks~\cite{CNN} (CNN), Transformers only use a mechanism called self-attention to capture dependencies between the input and output. 
The current state-of-the-art NLP models for programming languages such as CodeT5~\cite{CodeT5}, CodeBERT~\cite{CodeBERT} and PolyGlotCodeBERT~\cite{PolyglotCodeBERT} are all based on the Transformer architecture~\cite{Transformers}.
    
\subsection{Transfer Learning}
Pre-trained Transformers-based language models, such as RoBERTa~\cite{roberta}, CodeBERT~\cite{CodeBERT} and CodeT5~\cite{CodeT5} utilise a pre-train then fine-tune paradigm. The bespoke paradigm was initially introduced by~\citeauthor{BERT}. In this paradigm, the models are first trained in an unsupervised manner on a large unlabelled dataset. These pre-trained models can then be fine-tuned to perform a more specialised task, such as summarisation.
Transfer learning uses the knowledge that is obtained in one task to solve a different task. It allows the creation of general models that are trained once on massive datasets. These general models, which contain general domain knowledge can then be fine-tuned for a specific downstream task. This approach is quicker and requires less training data than training a model on the downstream task from scratch~\cite{BERT}.

\section{CAPYBARA Dataset}
\label{dataset}
We require a dataset of decompiled functions labelled with a descriptive summary to create and assess our solution. This dataset should be relatively large to suit the `data-hungry' nature of deep-learning models. Furthermore, the dataset needs to feature a diverse set of data representative of our solution's actual real-life use case. 
    
\subsection{Data Collection}
\begin{figure}
    \centering
    \includegraphics[width=\linewidth]{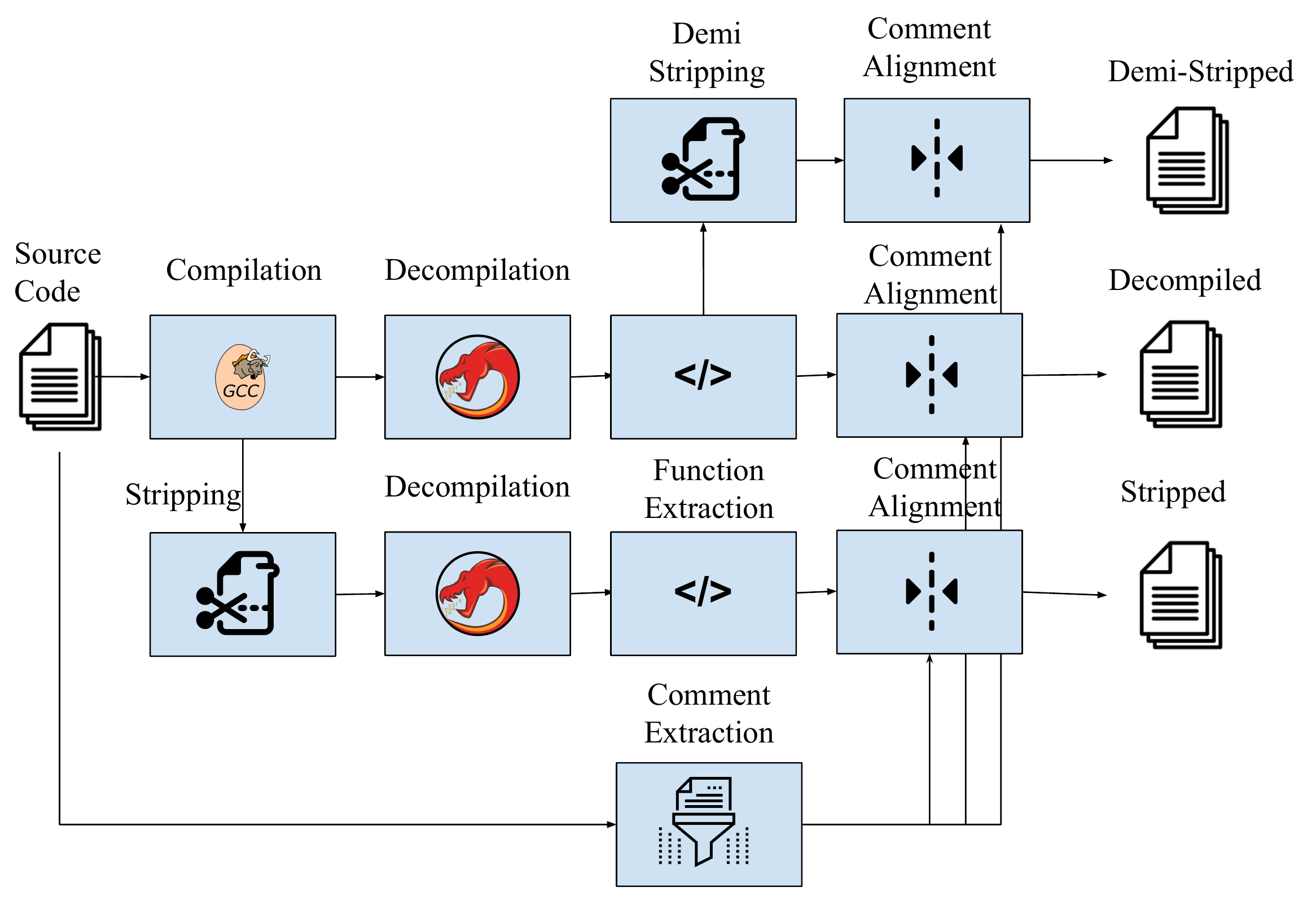}
    \caption{Data Collection Pipeline}
    \label{fig:dataCollection}
\end{figure}

To create such a large and diverse dataset we made use of BinSwarm~\cite{FUNCRE}, an existing dataset of aligned decompiled and stripped decompiled functions\footnote{BinSwarm: \url{https://hub.docker.com/r/binswarm/cbuilds}}.
BinSwarm collects C-based projects from Github. The projects are filtered to only include those that are actively being developed, using Travis CI and built for Ubuntu Linux. The projects are built using Docker. The resulting binaries are then copied and stripped, and both the stripped and unstripped binaries are decompiled using Ghidra. The functions are extracted from the stripped and unstripped decompiled code and aligned with the source code. The BinSwarm dataset only contains aligned tuples of source code and (stripped-) decompiled functions.
We extract documentation from the original source code files to add descriptive comments to this dataset. To that end, we depend on the documentation included in the source code by the original authors in the form of single and multiline comments. We locate the functions in the unbuilt project files and align the decompiled functions with the comments in the source code using srcML\footnote{srcML: \url{https://www.srcml.org/}} to extract any documentation located directly before a function signature. A high-level overview of the entire process is shown in Figure \ref{fig:dataCollection}.

A function's documentation often also contains other details besides the descriptive summary. We found that C projects do not follow a single documentation standard. For example, Javadoc for Java has a short one-line description or summary for each method at the beginning of the multiline comment block. In C, there is no singular documentation standard, so there might not be a single-line summary, and we will need to locate it in the comment block automatically.

\paragraph{Summary Extraction Rules}
We observe that the majority of single-line data are descriptive summaries, so we extract the first sentence. We identify many documentation styles in our multi-line data, we define some automated rules to extract summaries from the documentation:
\begin{itemize}
    \item{\textbf{@brief or @purpose:}} If the documentation contains a `@brief' or `@purpose' tag, we extract the first sentence after the tag. The `brief` tag is part of the Doxygen documentation standard\footnote{Doxygen:\url{https://doxygen.nl/manual/docblocks.html}}, an example is shown in Figure~\ref{fig:docExample}\footnote{jeanthom/DirtyJTAG:rcc\_set\_mco:\url{https://gitlab.com/insane-adding-machines/unicore-mx/-/blob/master/lib/stm32/common/rcc_common_all.c\#L192}}. 
    \item{\textbf{Description:}} If the documentation contains a line with `Description:`, we extract the following sentence.
    \item{\textbf{@param or @v:}} Documentation that contains an `@v' or `@param' tag, usually has a summary in the sentence before the tag. We extract that sentence.
\end{itemize}

\begin{figure}
    \centering
        \begin{minted}[breaklines,escapeinside=||]{C}
/** |\textcolor{blue}{\textbf{@brief}}| Select the source of Microcontroller Clock Output
 * Exact sources available depend on your target.
 * On devices with multiple MCO pins, this function controls MCO1
 * |\textcolor{codegreen}{\textbf{@param[in]}}| mcosrc the unshifted source bits
 */
        \end{minted}
    \caption{Example of documentation from jeanthom/ DirtyJTAG: rcc\_set\_mco}
    \label{fig:docExample}
\end{figure}
\paragraph{Filtering Rules}
To improve the quality of the dataset we filter out samples based on the rules used by the CodeSearchNet dataset~\cite{Codesearchnet} included in the CodeXGlue benchmark for the summarisation task~\cite{CodeXGlue}:

\begin{itemize}
    \item{\textbf{Documentation length:}}
    We remove any summaries that are too long or too short and remove anything shorter than 3 or longer than 256 tokens.
    \item{\textbf{Special tokens:}}
    We follow the example of the CodeSearchNet~\cite{Codesearchnet} and remove all documentation that contains special tokens. We scan for web tokens (like `http://'), HTML tokens (like `\textless head\textgreater'), paths (like `C://Users/..'), since this documentation usually refers to external resources. We additionally filter any developer tokens (like `FIXME:'), as these documents do not provide meaningful information about the function itself, but contain comments about the development process.
    \item{\textbf{Language:}}
    We filter out any documentation that was not written in English using the FastText language identification algorithm~\cite{fastText}. Around 92.19\% of the documentation is in English. 
    \item{\textbf{Empty documentation:}}
    We find that a large number of functions did not have any documentation associated with them at all. We simply remove these samples from the dataset.
     \item{\textbf{Abstract Syntax Tree:}}
    The authors of the CodeSearchNet dataset~\cite{Codesearchnet} 
    additionally, remove any samples that do not parse into an AST. We choose to omit this step since all of our samples have been successfully compiled and have thus at one point been parsed into an AST by the compiler.
\end{itemize}
\subsection{Dataset Preparation}
\paragraph{Synthesis of Demi-stripped Code}
From the dataset of decompiled functions, we also create another dataset. We emulate the process of stripping by removing all the identifiers from the decompiled code and replacing them with placeholders. For clarity, we call this demi-stripped data. Like the stripped dataset, the identifiers are all removed, but this is only done after the decompilation process. The decompiler still had access to the identifiers and could use the symbol table during decompilation. Most importantly, this demi-stripped dataset still has the same structure and control flow as the unstripped decompiled dataset and avoids any decompilation issues arising from stripping.

\paragraph{Data Split}
The dataset is split into a train, test and validation set. These sets constitute approximately, 80\%, 10\% and 10\% ~\cite{recommend_summarization} of the complete dataset. As recommended by~\citeauthor{evaluationSummarization} and~\citeauthor{recommend_summarization}, we prevent leakage of vocabulary and code patterns between the sets, by sampling the sets in a cross-project manner~\cite{evaluationSummarization, recommend_summarization}. This means that an entire project gets assigned to one of the sets, and functions from the same project cannot be assigned to different sets. The projects in the test and validation set are the same across all datasets.

\paragraph{Duplication}
Large corpora of code, like the corpus gathered by BinSwarm, tend to have a high degree of duplication~\cite{recommend_summarization}. As a result, snippets of code that are relatively unchanged appear in multiple parts of the corpus. This can be in the form of copied, generic or auto-generated functions. These functions will appear in multiple repositories and might be duplicated across the training and testing data.
Besides exact duplicates, near-duplicates can also occur. Near-duplicates differ in a few minor aspects like additional code comments or different function names. While removing exact duplicates is relatively fast and straightforward, removing near-duplicates is much more challenging and computationally intensive~\cite{allamanis_adverse}.
The issue with code duplication in classical code summarisation is that the models and tools are supposed to be used to generate summaries for new and unseen code. The evaluation metrics should therefore measure the generalisation of the tool on new samples~\cite{allamanis_adverse}. Duplicates and near-duplicates are not defined as new samples. A user of such a tool could simply look these samples up. Furthermore, large, high-capacity models like CodeT5 with 220M~\cite{CodeT5} or CodeBERT with 128M~\cite{CodeBERT} parameters, have a large capacity to memorise duplicated code~\cite{allamanis_adverse}.

However, the use case outlined in this work is more akin to deobfuscation. As explained by~\citeauthor{allamanis_adverse}, deobfuscation could be a use case where duplicates are valid and part of the true distribution of the problem~\cite{allamanis_adverse}. Compiled code contains a lot of duplicate code, and understanding this code is still difficult and essential for understanding the binary. While regular source code allows the reader to look up code snippets, decompiled binaries have an additional obfuscation applied. We, therefore, focus on the model's performance on code with duplicates as we believe duplicates to be part of the true distribution of the data, but we also report the deduplicated results.

\subsection{Dataset Properties}
Table~\ref{tab:dataset} shows the size of the processed dataset. Of the 2.1M aligned decompiled functions, we extract documentation for 215k of them, and we found that the majority of samples, 1.5M did not have any documentation at all. Furthermore, BinSwarm only provided us with 415k aligned stripped samples, and we can extract documentation for only 14k of these samples.
\begin{table}[htb]
    \centering
    \begin{tabular}{l|rr}
    \noalign{\smallskip}\toprule
    Dataset      & Including duplicates & Deduplicated \\ 
    \noalign{\smallskip}\bottomrule\noalign{\smallskip}
    C/Demi/Decom & 214,587              & 79,673       \\
    Stripped     & 14,245               & 7,826       \\
    \bottomrule
    \end{tabular}
    \caption{Number of functions in dataset}
    \label{tab:dataset}
\end{table}

The vast majority of documentation is in the form of multi-line comments as opposed to single-line or double-slash comments. We found that the documentation and comments had a mean length of 42.60 and 8.14 tokens, respectively.

\begin{figure}
    \centering
    \includegraphics[width=0.9\linewidth]{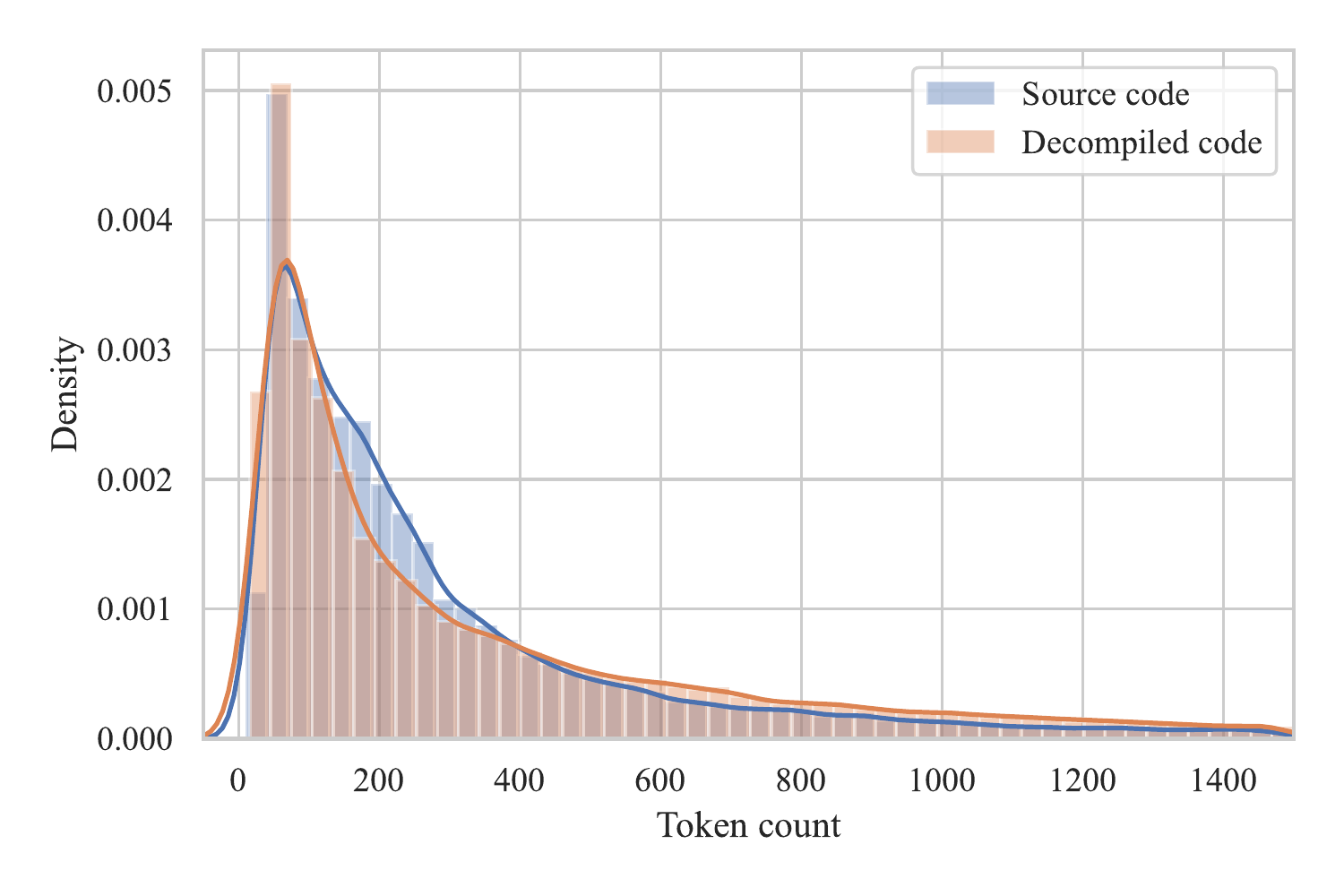}
    \caption{Tokens in source C and decompiled code}
    \label{fig:codeTokens}
\end{figure}

Figure~\ref{fig:codeTokens} shows the distribution of the number of tokens in source code and decompiled code. The source and decompiled code have a mean length of 399 and 779 tokens, respectively. Decompiled code also has close to double the LOC of source code, with means of 30.77 and 53.42 lines for source and decompiled, respectively.

\begin{figure}
    \centering
    \includegraphics[width=0.9\linewidth]{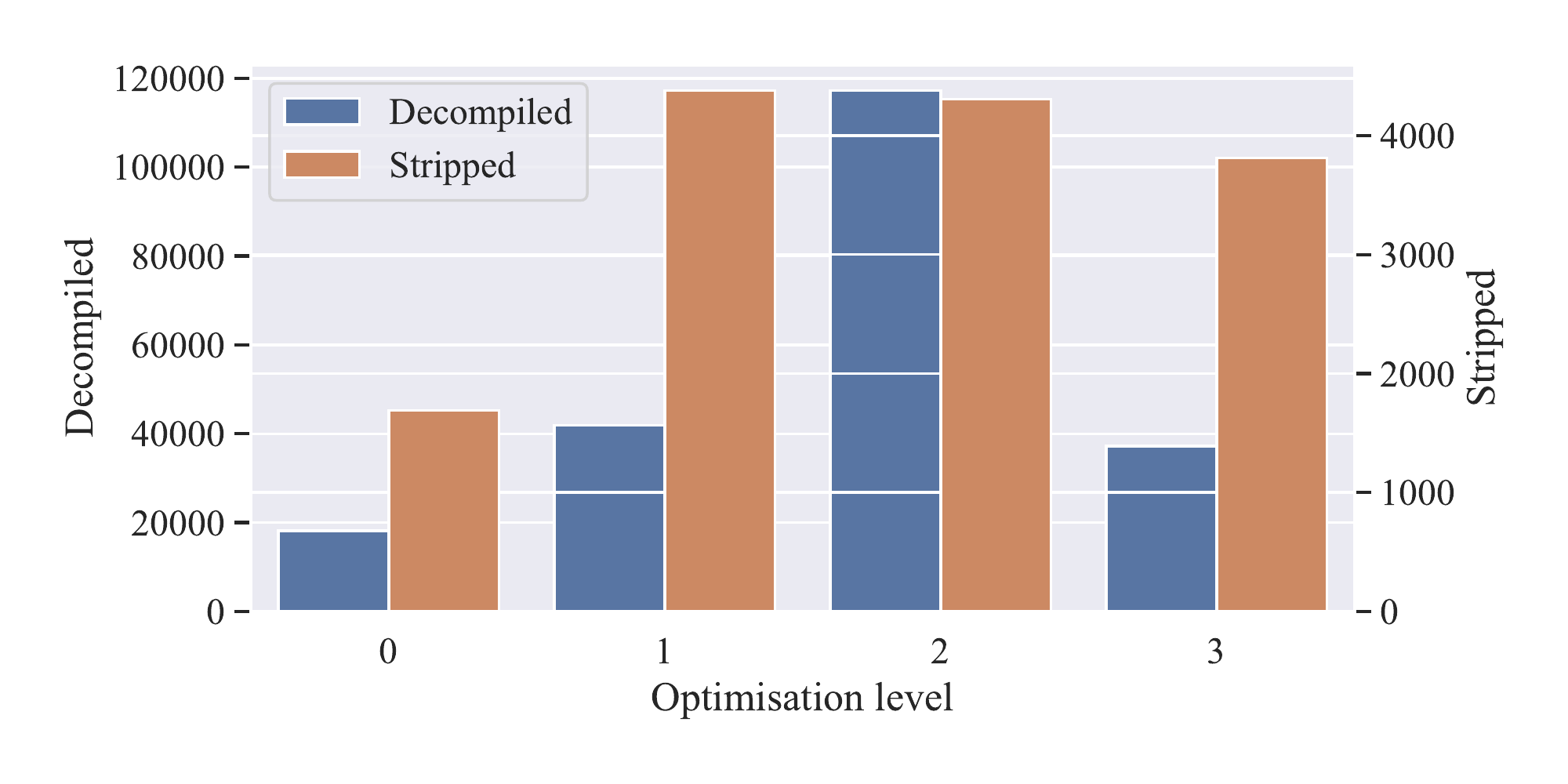}
    \caption{Distribution of optimisation levels in decompiled (left) and stripped (right)}
    \label{fig:optDistribution}
\end{figure}

The majority of decompiled functions are compiled with optimisation level -O2, with a similar number of -O1 and -O3 samples and relatively few -O0 samples. Stripped data has a very even distribution of optimisation levels, with only -O0 having significantly fewer samples. Note that there are more optimisation levels than shown in Figure~\ref{fig:optDistribution}, for brevity the different levels are grouped into their base optimisation level. -Oa is grouped with -O0, -Of and -Og are grouped with -O1, -Os is grouped with -O2. We also observe some samples with an optimisation level higher than -O3 (-O8 and -O7), as specified by the GCC documentation, these levels are equivalent to -O3\footnote{GCC optimisation levels: \url{https://gcc.gnu.org/onlinedocs/gcc-4.4.2/gcc/Optimize-Options.html\#Optimize-Options}}. 

\section{BinT5}
\label{model}
We select CodeT5~\cite{CodeT5} as the base-model for our experiments since it is the highest-scoring publicly-available model on the CodeXGLUE~\cite{CodeXGlue} Code Summarisation benchmark\footnote{CodeXGLUE benchmark: \url{https://microsoft.github.io/CodeXGLUE/}}. CodeT5 is a programming language model built on the T5 (Text-to-text Transfer Transformer) architecture~\cite{T5} and pre-trained on a mix of supervised and unsupervised tasks. CodeT5 employs an encoder-decoder architecture.
In contrast to other models, CodeT5 is trained using both unimodal (PL only) and bimodal (NL-to-PL) tasks in eight programming languages. This bimodal training allows CodeT5 to perform strong cross-modal tasks such as code summarisation and code generation (PL-to-NL). Many other models only use the data and languages included in the CodeXGlue dataset~\cite{CodeBERT, PolyglotCodeBERT,CodeXGlue}, while CodeT5 also uses a mined dataset of C and C++ code for its pre-training objectives~\cite{CodeT5}. The inclusion of C training data should help the model with the CAPYBARA dataset. There could be some overlap in the training data between CAPYBARA and the dataset used by~\citeauthor{CodeT5} which would cause leakage, we address these concerns in Section~\ref{discussion}.

\begin{figure}
    \centering
    \includegraphics[width=0.9\linewidth]{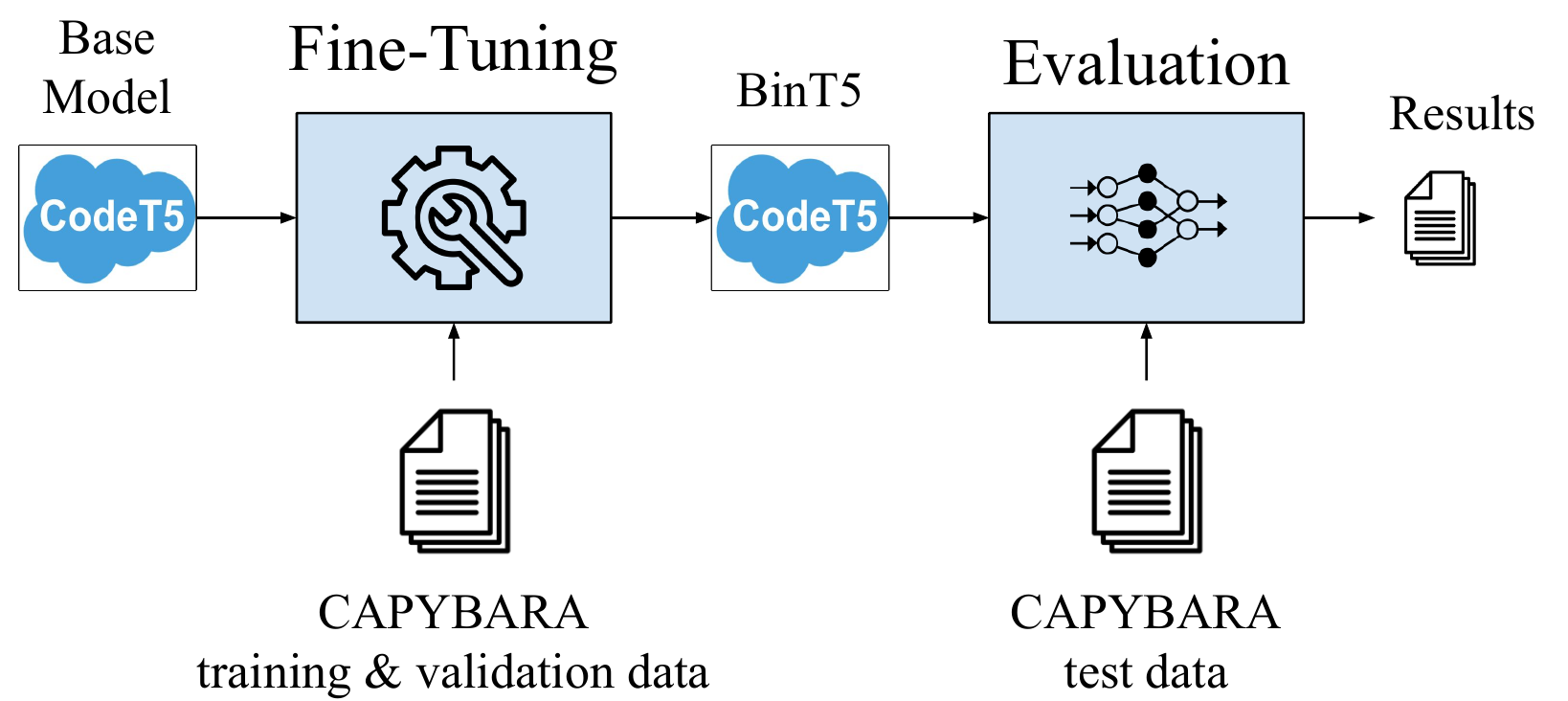}
    \caption{BinT5 fine-tuning pipeline}
    \label{fig:fine-tuning}
\end{figure}

CodeT5 also utilises the transfer learning paradigm, which allows us to train the model with relatively little data. In this case, we make use of the CodeT5-base model, which was trained on mixed upstream tasks by the authors~\cite{CodeT5}. We fine-tune this model on the code summarization task on CAPYBARA. An overview of how we applied the model to create BinT5 is provided in Figure~\ref{fig:fine-tuning}. 

\section{Experimental Setup}
\label{expSetup}
To assess the effectiveness of our approach, we first evaluate the performance of the model, we then identify the aspects of the data that make this task inherently difficult, and we finally investigate aspects of the datasets and their influence on the complexity of the task. 
\subsection{Research Questions}
In the context of the study, we thereby formulate the Research Questions (RQ) as follows.
\begin{itemize}
    \item[RQ1:] \textit{How effective are fine-tuned Transformer-based models at decompiled code summarisation?} To investigate the application of existing models to binaries using CAPYBARA, we set a baseline by training a model on the code summarisation task on the source C-code dataset. We then train a summarisation model on both the decompiled and the stripped dataset. We use the evaluation metrics to compare the performance of the different models.
    \item[RQ2:] \textit{Which aspects of the input contribute most to model performance?} We investigate which aspects of decompiled code increase the difficulty of the task. We, therefore, look at the impact of the symbol table on decompilation, for this, we fine-tune a model on the demi-stripped dataset and compare it to the other models. We also investigate the importance of the function name by removing \textit{just} the function name from the decompiled code. Furthermore, we investigate the impact of the optimisation level by exploring the performance per optimisation level.
    \item[RQ3:] \textit{What is the impact of dataset properties on model performance?} We finally investigate how the construction of CAPYBARA influences the models. To answer the final research question we remove the duplicates from the datasets and retrain the models, after which we compare the performance to the baselines. Furthermore, we investigate the impact of dataset size, by incrementally reducing the size of the training sets.
\end{itemize}
\subsection{Baselines}
To first establish a performance baseline, we train a CodeT5-base model on the summarisation task on source C. Note that only samples which are aligned with decompiled code are included in the source C dataset. The baseline is used to compare the decompiled C, stripped decompiled C and the demi-stripped datasets to the source code.
\subsection{Evaluation Metrics}
We evaluate the performance between the reference summary from CAPYBARA and the candidate summary produced by BinT5 using the EM, BLEU-4~\cite{BLEU}, ROUGE-L~\cite{Rouge} and, METEOR~\cite{Meteor} metrics. 
\paragraph{Exact Match (EM)}
The simplest metric is the EM which scores a prediction one if it matches its reference exactly and zero otherwise. 
\paragraph{BLEU-4}
The most widely used metric in the code summarisation task is the Bilingual Evaluation Understudy Score (BLEU)~\cite{evaluationSummarization}. BLEU-4 produces a percentage number between 0 and 100, which defines the similarity between a candidate and a set of reference sentences. 
BLEU-4 calculates the cumulative 4-gram precision scores, the number of matching 4-grams divided by the total number of 4-grams in the candidate sentence~\cite{BLEU}. The unigrams and bigrams account for the adequacy of the candidate while the longer three and 4-grams account for fluency. To prevent short sentences the result is multiplied by a brevity penalty as well. A smoothing function is applied to prevent sequences with no matching 4-grams to score zero~\cite{Smoothing}.  While~\citeauthor{evaluationSummarization} recommend BLEU-4 with smoothing method 4~\cite{evaluationSummarization}, we opted to use the Moses~\cite{Moses} implementation of BLEU-4 which uses smoothing method 2 since this is also utilised by CodeSearchNet, CodeXGlue and CodeT5~\cite{Codesearchnet, CodeT5, CodeXGlue}.
\paragraph{ROUGE-L}
ROUGE or Recall-Oriented Understudy for Gisting Evaluation, is a package which includes several metrics, the most popular among them is ROUGE-L~\cite{Rouge}. ROUGE-L is more recall oriented than BLEU-4. ROUGE-L simply finds the longest common subsequence (LCS) between the reference and the candidate. Note that the words do not need to be consecutive but they have to be in order. 
\paragraph{METEOR}
METEOR or Metric for Evaluation for Translation with Explicit Ordering~\cite{Meteor} uses word lists and stemming to also take synonyms into account and calculates the harmonic mean of the unigram precision and recall. Similar to ROUGE-L, METEOR is more recall-focused. METEOR has a higher correlation with human judgement than BLEU-4~\cite{recommend_summarization} at the sentence level.

\subsection{Data deduplication}
To create a deduplicated version of the CAPYBARA dataset we make use of a fork\footnote{Near Duplicate Code Detector: \url{https://github.com/SERG-Delft/near-duplicate-code-remover}} of the near-duplicate-code-detector~\cite{allamanis_adverse}. We use this tool to compare all the datasets' functions and find clusters of near-duplicate functions. We randomly select one function per cluster and discard the rest from the dataset. We use the standard tool configuration as recommended by~\citeauthor{allamanis_adverse}. 
Of the removed duplicates, we observe that a relatively large number originates from common libraries, such as SQLite\footnote{SQLite: \url{https://www.sqlite.org/index.html}}, that are packaged with binary programs. Thus a certain amount of duplication is also likely to occur ``in the wild".

\subsection{Configuration}
We process and visualise the data with Pandas $1.4.3$ and Ghidra $10.0.4$\footnote{It is not recommended to use Ghidra versions before 10.1 since these versions have not been patched against a Log4J RCE}. FastText $1.0.3$ with the largest lid.176.bin model is used to detect languages. We train the model using Transformers version $4.16.2$ running on Torch $1.9.0$+cu$111$ in the nvidia/cuda:11.4.0-base docker container image. We share a Docker image with all the libraries required to run BinT5 pre-installed on DockerHub\footnote{BinT5 Docker Image: \url{https://hub.docker.com/r/aalkaswan/bint5/tags}}.

A grid search of the optimal settings was infeasible from a time perspective, so we performed training mainly using the recommended settings from the CodeT5-base model~\cite{CodeT5}. We double the source length for the decompiled, stripped, and demi-stripped code to $512$ tokens instead of the standard $256$ tokens used for the source code to compensate for the fact that the average length of decompiled code is almost twice as long as the source code. 
We trained the model on a machine with an NVIDIA GeForce RTX3080 with $10$GB of VRAM and an AMD Ryzen Threadripper 3990X 64-Core Processor with $192$GB of RAM running Ubuntu $20.04.4$ LTS. The GPU is running Nvidia driver version $510.60.02$ with Cuda $11.6$. The authors of CodeT5 used an NVIDIA A100 GPU with $40$GB of VRAM for fine-tuning~\cite{CodeT5}. To compensate for the lack of memory, we reduced the batch size to $2$, which was the maximum length that could still fit in the VRAM, we increase the `gradient\_accumulation\_steps' to $24$ to still achieve the effective standard batch size of $48$. %(see Table:~\ref{tab:modelSettings}).

% \begin{table}
%     \centering
%     \begin{tabular}{lll}
%         \toprule
%                         &CodeT5-base & BinT5                    \\ \midrule
%         Source length   & 256 Tokens & \textbf{256/512 Tokens} \\
%         Target length   & 128 Tokens & 128 Tokens              \\  
%         Max epochs      & 15         & 15                      \\  
%         Patience        & 2          & 2                       \\ 
%         Batch Size      & 48         & \textbf{2*24}           \\ 
%         Vocabulary Size & 32100      & 32100   \\
%         \bottomrule
%     \end{tabular}
%     \caption{Model configuration}
%     \label{tab:modelSettings}
% \end{table}

\section{Results}
\label{results}
We present the results of our experiments to answer the research questions, results are grouped per research question. The metrics are calculated for each sample from the test set, and the average scores are presented.
%%%%%%%%%%%%%%%%%%%%%%%%%%%%%%%%%%%%%%%%%%%%%%%%%%%%%%%%
\subsection{RQ1: Model Effectiveness}
The performance of the CodeT5-base model on each of the datasets is presented in table~\ref{tab:duplicated}. 
\begin{table}
    \centering
    \begin{tabular}{lllll} 
        \noalign{\smallskip}\toprule
                 & BLEU-4 & EM & METEOR & ROUGE-L  \\ 
        \cmidrule{2-5}
        C        & 60.83  & 52.19 & 65.33  & 66.51           \\
        DecomC   & 58.82  & 48.92 & 63.14  & 64.51      \\
        Stripped & 11.26  & 1.85  & 14.50   & 17.25    \\
    \noalign{\smallskip}\bottomrule
\end{tabular}
    \caption{Result of fine-tuning CodeT5-base on mined datasets}
    \label{tab:duplicated}
\end{table}

We found that the decompiled code model generally produced good summaries, evidenced by the BLEU-4 score of 58.82, which is slightly lower than the baseline set by the source code. The stripped model mainly produced unusable summaries, as evidenced by the BLEU-4 score of 11. The high EM score could be an indication of a high duplication factor.

Initial experiments with GraphCodeBERT~\cite{graphCodeBERT} and PolyglotGraphCodeBERT~\cite{PolyglotCodeBERT} base models fine-tuned on CAPYBARA show performance around 5 and 3 BLEU-4 lower, respectively. This is a relatively small difference, especially considering the model size. This shows that the performance of BinT5 does not heavily depend on the additional pre-training on C and C\# performed by~\citeauthor{CodeT5}. Furthermore, this result shows that it is improbable that significant dataset leakage has taken place.

We found a relatively large difference between the number of recovered decompiled and stripped decompiled functions. This can likely be attributed to the fact that Ghidra struggles a lot more with recovering stripped functions. Recall that the symbol table commonly contains information regarding the location and name of functions. When this table is dropped, the start- and endpoints of functions are hard to infer by automatic tools, especially since many functions get inlined, and \code{JUMP} instructions replace \code{CALL} instructions. Aside from difficulties in demarcating functions, it is also difficult to align the associated source code function with the decompiled function. With unstripped code, the function name remains, meaning the functions can be aligned using the name. We attempted to utilise an existing solution by~\citeauthor{FunctionBoundaryDetection} called Jima~\cite{FunctionBoundaryDetection} to find function boundaries. Jima is the current state-of-the-art tool for function boundary detection in stripped binaries. The tool is implemented as a plugin for Ghidra, but in our experiments, we find no statistical difference between the base performance of Ghidra and Jima on our dataset. The difficulties in extracting stripped functions, make training and applying a model to stripped binaries challenging.

%%%%%%%%%%%%%%%%%%%%%%%%%%%%%%%%%%%%%
\subsection{RQ2: Input Properties}
\begin{table}
    \centering
    \begin{tabular}{lllll}
    \noalign{\smallskip}\toprule
             & BLEU-4 & \multicolumn{1}{c}{EM} & \multicolumn{1}{c}{METEOR} & ROUGE-L  \\ 
    \cmidrule{2-5}
    DecomC    & 58.82  & 48.92 & 58.4   & 60.32                                               \\ 
    Demi      & 44.21  & 35.10 & 47.89  & 49.59                                               \\
    NoFunName & 46.99  & 37.12 & 45.92  & 48.07  \\
    \noalign{\smallskip}\bottomrule
    \end{tabular}
    \caption{Result of fine-tuning CodeT5-base on synthetic data}
    \label{tab:demi}
\end{table}

As can be observed in Table \ref{tab:demi}, the summaries produced by the demi-stripped model were substantially worse than the decompiled model, but most were still very usable, evident by the BLEU-4 score above 44. Just removing the function name gave quite similar results to demi-stripping. 
We find that the loss of identifiers significantly lowers the performance of the model, but stripped code also suffers from decompilation faults, which seem to have a much larger impact on the model performance. Hence, the performance of BinT5 on demi-stripped code can be viewed as more representative of the actual model and not impacted by faults introduced by Ghidra.

\begin{table}
    \centering
    \begin{tabular}{lllll}
    \noalign{\smallskip}\toprule
    Opt level & BLEU-4 & EM & METEOR & ROUGE-L  \\ 
             \cmidrule{2-5}
    -O0      & 72.88 & 34.18 & 73.19  &  74.84  \\ 
    -O1      & 50.30 & 59.84 & 55.36  &  54.84  \\ 
    -O2      & 62.31 & 46.23 & 64.50  &  66.05  \\ 
    -O3      & 54.68 & 54.99 & 58.25  &  59.28   \\
    \noalign{\smallskip}\bottomrule
    \end{tabular}
    \caption{Average BLEU-4 score of decompiled code per optimisation level}
    \label{fig:opt}
\end{table}
Table~\ref{fig:opt} shows the average score per optimisation level. We can observe that -O0 and -O2 perform better than -O1 and -O3. Recall that -O0 is completely unoptimised, and that the vast majority of our decompiled dataset is compiled with -O2, which would explain why those optimisation levels perform better.

%%%%%%%%%%%%%%%%%%%%%%%%%%%%%%%%%%%
\subsection{RQ3: Dataset Properties}
The performance of the base model on each of the deduplicated datasets is presented in table~\ref{tab:deduplicated}:
\begin{table}[htb]
    \centering
    \begin{tabular}{lllll|l}
    \noalign{\smallskip}\toprule
             & BLEU-4 & EM    & METEOR & ROUGE-L & $\Delta$BLEU-4 \\ 
             \cmidrule{2-6}
    C        & 45.86  & 32.87 & 46.06  & 47.53   & 14.97          \\ 
    DecomC   & 42.48  & 28.08 & 25.23  & 27.66   & 16.34          \\ 
    Demi     & 25.38  & 14.51 & 42.47  & 44.47   & 18.83          \\ 
    Stripped & 7.19   & 0.00  & 4.75   & 5.50    & 4.07          \\
    \noalign{\smallskip}\bottomrule
    \end{tabular}
    \caption{Result of fine-tuning CodeT5-base on the deduplicated datasets and the difference with the baseline}
    \label{tab:deduplicated}
\end{table}
    
We find that the influence of deduplication on our model's performance is relatively small on source code, at only 24\%. Duplicates have a relatively large impact on the decompiled (28\%) and demi-stripped (43\%) code. Deduplication also greatly decreases the EM rate across the board.
Duplicates have a relatively large impact on performance, but even with the duplicates removed the model still produces many high-quality summaries. 
%As noted previously in Chapter~\ref{methodology}, duplicates are part of the problem space. We, therefore, consider them in the other experiments. 
The experiments on deduplication show that the model seems to have a deeper understanding of the data and is not simply reproducing previously seen samples.

\begin{figure}
    \centering
    \includegraphics[width=0.9\linewidth]{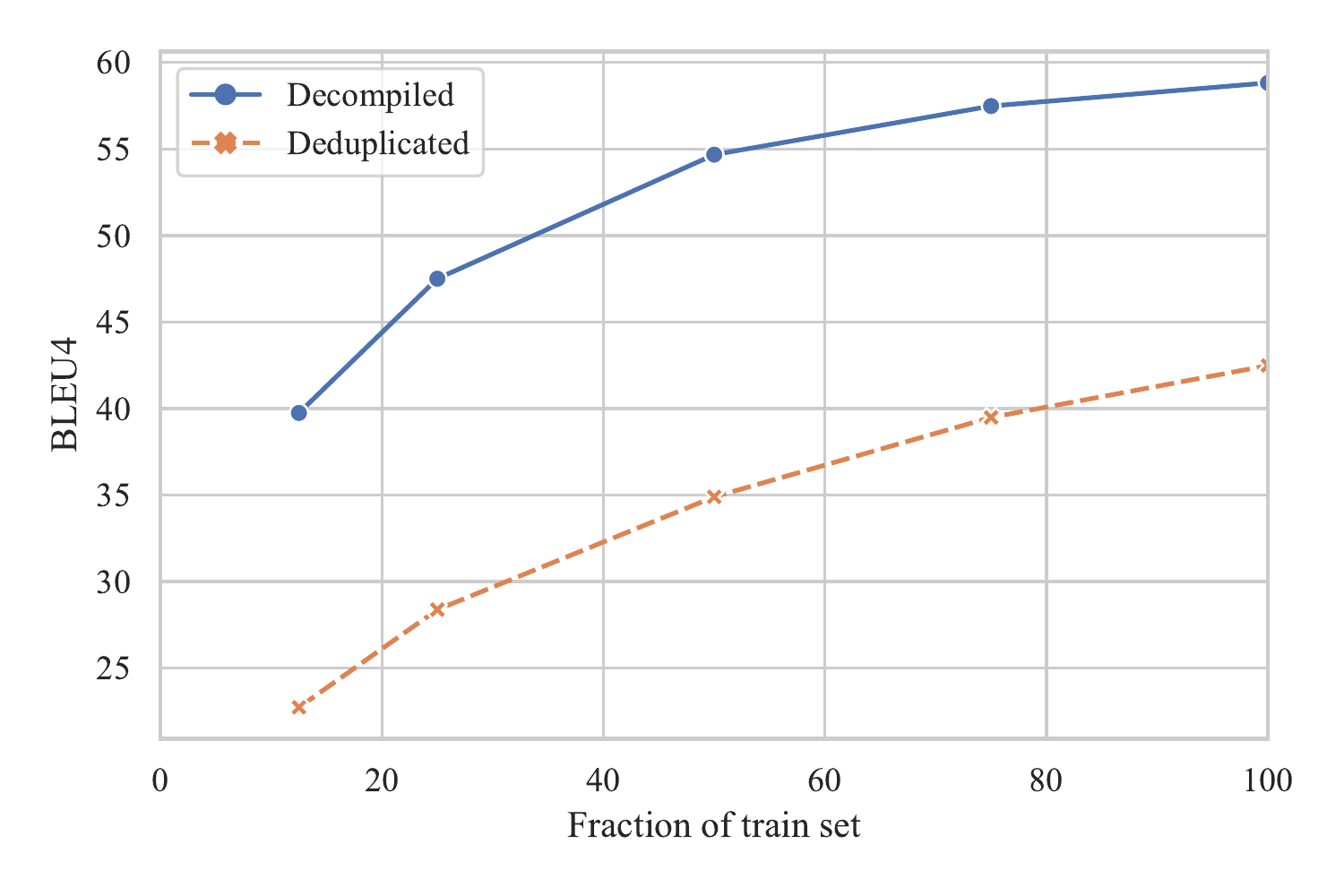}
    \caption{BLEU-4 per trainset size for decompiled code and deduplicated decompiled code}
    \label{fig:trainSize}
\end{figure}

As can be seen in Figure~\ref{fig:trainSize}, the dataset size does not have much of an impact, the model can be trained with half or a quarter of the training samples without suffering a considerable hit to performance. This could be attributed to the high duplication factor of our dataset. It could also be because the model was already pre-trained well by~\citeauthor{CodeT5} and requires very little data for fine-tuning. This is a testament to the relative ease with which these models could be extended to decompiled code.

% \paragraph{Filtering Policies}

We also performed experiments where we did not apply the filtering rules provided by CodeXGlue and where we always mined the first sentence of any type of documentation. While we were able to collect around 480K decompiled samples, the model performed substantially worse, only scoring 36.97 and 33.26 BLEU-4 on C and decompiled code, respectively. These results show that the dataset quality also heavily impacts the model performance.  

\section{Discussion}
\label{discussion}
In the previous section, we found that BinT5 shows considerable performance for decompiled code and demi-stripped code on both regular as well as deduplicated data. While this is a promising result, we conduct a small investigation of the decompiled samples. We will put our observations on identifiers into the context of the extreme summarisation task. Based on this we discuss the implications of our work. Finally, we will close this section by discussing the threats to validity.

\subsection{Exploration of Results}
To explore the results of BinT5 we pick 25 high and 25 low-scoring samples from the test set of the deduplicated decompiled dataset. High samples have a BLEU-4 score higher than 75 while low-scoring samples have a score lower than 25.

\paragraph{High Samples} With the high-performing samples BinT5 tends to produce summaries which are very close to the references. For instance, BinT5 produced \code{Print description of a datatype in XML} against the baseline \code{Dump description of a datatype in XML}. Of the 25 high-scoring samples we found that all have counterparts with a similar function summary in the training set. These functions also tend to have similar names, but their decompiled function body was significantly different, which is likely why deduplication didn't remove these functions.

\paragraph{Low Samples} From the low-performing samples we observe that many summaries produced by BinT5 are semantically very similar to the reference. For instance, the function \code{vl\_set\_simd\_enabled}\footnote{Colmap/Colmap:vl\_set\_simd\_enabled: \url{https://github.com/colmap/colmap/blob/87b3aa325bd8e5fb913788e29e9ac1e085e28b67/lib/VLFeat/generic.c\#L1070}}, has the reference \code{Toggle usage of SIMD instructions} while BinT5 produced \code{Enable or Disable the Simd Channel}. This sample scores a BLEU-4 score of 0.0, because of the limitations around the BLEU-4 metric, while for a human evaluator the output is still very usable. 
Similarly, for some samples, BinT5 produces shorter summaries containing shorthands. The reference \code{Check if the given nickname is blocked for "normal client" use} against \code{Check whether nick is blocked}, also scores poorly. 
Of the 25 low-scoring samples we observe that around 11 are semantically similar to the reference and likely very useful for understanding the function.

\subsection{Identifiers and Extreme Summarisation}
We find a relatively small difference in performance between source code and decompiled code. This indicates that in-function comments and variable names are relatively unimportant for the model performance. Although~\citeauthor{PolyglotCodeBERT} observed that identifiers might be more important than syntax in the code-summarisation task~\cite{PolyglotCodeBERT}, we can further conclude that the function name is explicitly essential for model performance. Removing just the function name from the decompiled samples, as opposed to removing all identifiers in demi-stripping, results in slightly higher performance than demi-stripped code, which indicates a very high dependence on the name of the function in the code summarisation task, which is a logical finding in the context of the extreme code summarisation task. 

The extreme code summarisation task, as proposed by~\citeauthor{ExtremeSummarization} aims to reproduce the function name given a function body \cite{ExtremeSummarization, PolyglotCodeBERT}. It is framed as a summarisation problem where the output is around 3 tokens in length, instead of the 10+ tokens that regular code summarisation targets. We found similar results when performing this task with our dataset, namely, high performance on regular decompiled code (with function names removed) and low performance on stripped code.

A manual assessment of the stripped data shows that many of the aligned functions were not decompiled properly. We find that many functions are cut-off after a few instructions because the decompiler did not recover the full control flow. Other functions are missing side effects, like changes to global variables.

\subsection{Implications}
We propose a novel solution to aid reverse engineers in their work. If the application of NLP to binaries gets significantly better, and the limitations around stripping and other obfuscation techniques get resolved, it would have severe implications for the cybersecurity domain. On one hand, it could help malware analysts understand novel malware and its weaknesses quickly. Software can be analysed to find possible vulnerabilities and malicious payloads. Source code can be reconstructed for old binaries for which the source code is lost. But on the other hand, attackers can leverage these same methods to find and exploit vulnerabilities and lift intellectual property from binaries.

CAPYBARA itself could be used to create and assess neural decompilation,
to perform a deeper investigation into the extreme summarisation task,
or to simply train a code summarisation model on C code. 
CAPYBARA consists of a large corpus of C and decompiled C code, 
which could be used to pre-train language models, 
such that these models could support decompiled code out-of-the-box.

While our work focused on decompiled code, our observations show some limits of transformer-based models and their applicability to different data. Our dataset can help and inspire other researchers to improve upon our work.
We hope other researchers use this dataset to train and evaluate their own models. Furthermore, the process outlined in Chapter~\ref{dataset} could help others construct standardised datasets for other tasks and languages. 

\subsection{Threats to Validity}
\textbf{Internal Validity} questions if other factors could have affected the outcome. The training and evaluation data contains a significant amount of noise, either in the form of badly decompiled functions or incorrect documentation. We carefully collect and process the data, but we are unable to know to which extent the documentation matches the original code. While machine learning models (and specifically NLP models) should be able to handle noisy data, this might introduce some bias into the models.
CodeT5 was also pre-trained on a C and C\# dataset, this dataset is unpublished and we were unable to reach the authors. Some data leakage might have taken place, but as explained in Section \ref{results} it is unlikely that it had much of an impact. To prevent this threat from arising in any future studies, we make CAPYBARA publicly available.

\textbf{External Validity} refers to the generalisability of our results. This work only focuses on stripping and compiler optimisations as a means of resisting binary analysis, other techniques like control flow obfuscation and packing are also used to prevent reverse engineering. Other works focus on unpacking and deobfuscation, so we consider our work orthogonal to theirs. 
The data gathered for CAPYBARA were exclusively from open-source projects. Decompiling closed-source projects is explicitly forbidden by some EULAs and the lack of source code documentation makes it difficult to evaluate using reference summaries. However, reverse engineering open-source software is not very useful in practice, since the source code is readily available. Closed-source software might have different data distribution and will present other challenges like obfuscation.
Finally, only functions that decompile (Ghidra produces any output) and that are documented, are represented in CAPYBARA. This is most apparent in the stripped dataset, where we can only recover a small fraction of the total number of functions. A deeper investigation into new decompilation techniques for stripped code, specifically into the aspect of function boundary detection is left as future work.

\textbf{Construct Validity} relates to the adequacy of the theoretical constructs and the use of appropriate evaluation metrics. The leading metric in our evaluations does not capture semantic meaning. While BLEU-4 is the most popular metric for this task, its reliability has been called into question~\cite{CodeSumMetrics, SentenceBERT}. We, therefore, included other metrics, which do take semantics into account, in our evaluation.
Finally, our entire approach hinges on the assumption that function summaries, as they are used for source code, are useful for binary analysis. Whether or not this is actually the case, should be further investigated with a qualitative user study, this is left as future work.
    
\section{Related Work}
Binary reverse engineering and the use of NLP for software engineering are vast and active fields, so we select and discuss the closest state-of-the-art works in the field. We categorise the studies into identifier recovery and binary translation. Finally, we will discuss the open challenges and the relation of our own work to these challenges. 

\paragraph{Recovering Identifiers from Stripped Binaries}
\textbf{Debin}~\cite{Debin} aims to recover debug information from stripped binaries. The authors use a tree-based classification and a probabilistic graph-based model. All the variable names and types are jointly recovered using a maximum a posteriori probability inference. \textbf{VarBERT}~\cite{VarBERT} uses a Transformer-based NLP model for the task of variable name recovery. The authors pre-trained a BERT model which is then fine-tuned to predict the names and types from \textit{unstripped} binaries.

\textbf{FUNCRE}~\cite{FUNCRE} uses a pre-trained and fine-tuned ROBERTA~\cite{roberta} model to predict usages of inlined library functions. Recall that compilers with optimisations enabled can inline functions in the binary (Chapter~\ref{background}). The authors use indelible markers, which do not get destroyed by the compiler, to mark usages of library functions and to construct a dataset and train a model.
\paragraph{Binary Translation}
\textbf{Neutron} \cite{Neutron} frames decompilation as a neural machine translation problem and utilises an Attention-LSTM-based neural translation network to translate disassembled binaries back to C source code. The binaries are not stripped and do not have any optimisations enabled. The translations created by Neutron can contain syntax errors, so the authors apply regular expressions to create a tailor-made syntax checker. Neutron achieves high accuracy on the translation task, but only on unstripped and non-optimised code.

\paragraph{Our Novelty}
Several aspects have not been properly addressed and investigated. 
The application of code summarisation methods to decompiled code has not been addressed by any work at all. Furthermore, some works on binary code fail to take compiler optimisations into account~\cite{Neutron}. We, therefore, investigate the application of code summarisation methods to decompiled code and we enable compiler optimisations.
   
\section{Conclusion}
In this paper, we proposed a new automatic binary code summarisation task. With this new task, we also introduce CAPYBARA, a novel dataset to train and evaluate models on this task, with both mined as well as synthetic data. Paired with this dataset, we train BinT5, a Transformer-based code summarisation model to show the effectiveness of CAPYBARA. We used BinT5 to further explore the datasets, outlining the inherent difficulties in the data.

We found that while BinT5 shows considerable performance on regular decompiled code, but its performance is being hampered by the decompiler on stripped code, evidenced by BinT5s strong performance on demi-stripped code. Furthermore, we found that while duplicates have a large impact on the model, their presence is not paramount to the model's performance. Finally, we observe that BinT5 could be trained with just a fraction of the samples in CAPYBARA.

Our work has shown that a well-known and well-studied task from the source code domain~\cite{evaluationSummarization}, namely source code summarisation, can be applied to binary code. This is only one of the many different applications of NLP for code. Our paper constitutes the first step in the application of source code NLP methods to such tasks on binary code. 
\bibliographystyle{IEEEtranN}
\bibliography{references.bib}

\end{document}